\documentclass[%
 aip,
 jmp,%
 amsmath,amssymb,
 reprint,%
]{revtex4-1}

\usepackage{graphicx}
\usepackage{dcolumn}
\usepackage{bm}
\usepackage{url,graphicx,graphics,amssymb,amsmath,relsize,supertabular}
\usepackage[colorlinks=true,citecolor=blue,linkcolor=blue]{hyperref}%

\begin{document}


\title{Reducing the Efficiency Roll-Off in Organic Light-Emitting Diodes at High Currents under External Magnetic Fields}

\author{Afshin Shahalizad}
\author{Jean-Michel Nunzi}%
 \email{nunzijm@queensu.ca}

 \affiliation{Department of Physics, Engineering Physics \& Astronomy, Queen’s University, Kingston, Ontario K7L 3N6, Canada}
%

\date{\today}

\begin{abstract}
     Singlet-polaron and triplet-polaron annihilation mechanisms are the most detrimental exciton quenching processes that lower the efficiency of organic light-emitting diodes (OLEDs) at high current densities, causing so-called \emph{efficiency roll-off} in these devices. These exciton loss mechanisms are also the critical obstacles towards the realization of electrically pumped organic semiconductor lasers, which require very high current densities to reach threshold. Herein, under a relatively large external magnetic field, we demonstrate that the efficiency roll-off at high current densities in europium (Eu$^{3+}$)-based solution-processed OLEDs can be suppressed to some extent while the luminance is enhanced. We achieve this by reducing the F\"{o}rster-type exciton-polaron annihilation processes. Under the applied magnetic field, we show that manipulation of the polaron-spin and exciton dynamics lead to a quantitative roll-off suppression.
\end{abstract}

\maketitle

\section{INTRODUCTION}
\label{INTRODUCTION}
Despite the significant amount of attention that organic light-emitting diodes (OLEDs) have raised in the past decades, exciton-exciton and exciton-polaron annihilation mechanisms have remained the limiting factors for the achievable maximum efficiency and brightness required in displays and solid-state lighting, because of the so-called \emph{efficiency roll-off} phenomenon.\cite{1,2,3,4} Singlet-polaron (S-P) and triplet-polaron (T-P) annihilation processes are indeed the major efficiency loss processes under high current densities in OLEDs.\cite{5,6,7} Additionally, they hinder achieving electrically pumped continuous-wave lasing operation in OLEDs, which requires a minimum current density of 1 kA/cm$^2$.\cite{8,9} Several research groups have proposed various approaches to suppress the exciton-polaron annihilation.\cite{10,11,12,13,14} Nevertheless, these approaches may be rather inefficient, technically complicated, or costly to implement into practical applications. Therefore, development of effective, straightforward, inexpensive, and easy-to-implement techniques for reducing the exciton-polaron annihilation is of particular importance.\\
Previous studies have shown that applying external magnetic fields to fluorescent, phosphorescent, and thermally activated delayed fluorescent (TADF) OLEDs can enhance their performance.\cite{15,16,17,18,19,20,21,22,23} Various models have also been proposed to explain the observed magnetic field effects in these devices. These models mainly rely on the perturbation of the quantum-statistical 1:3 singlet-to-triplet exciton density ratio, \cite{15,16,17,18,19,20,21,22} or on the reduced spin-dependent reaction rate between triplet excitons and polarons under external magnetic fields.\cite{23,24} While all the previous studies have mainly focused on enhancing the electroluminescence (EL) intensity and magneto-conductivity of OLEDs, to our knowledge, there has been no report on the influence of external magnetic fields on suppression of the external quantum efficiency (EQE) roll-off to date.\\
In this paper, we experimentally investigate the influence of a relatively large external magnetic field on the EQE roll-off characteristics and luminance (brightness) of solution-processed OLEDs based on a Eu$^{3+}$-based lanthanide complex emitter. The complex has a very narrow-band red emission that makes it an ideal emitter for OLED displays. However, studies have shown that Eu$^{3+}$-based devices usually exhibit a low level of brightness and severe efficiency roll-off, making their applications rather challenging in practical OLED displays and, if mixed with blue and green emitters, in white OLEDs for solid-state lighting.\cite{25} Our results reveal that applying the magnetic field can enhance the luminance remarkably and reduce the EQE roll-off through mitigating the exciton-polaron annihilation at moderate and high current densities, owing to the manipulation of the polaron-spin precessions and subsequently perturbation of the singlet and triplet excited state dynamics.

\section{EXPERIMENTAL}
\label{EXPERIMENTAL}
We incorporated Eu(DBM)$_3$Phen lanthanide complex (with DBM: 1,3-diphenylpropane-1,3-dione and Phen: 1,10-phenonthroline ligands) (5 wt.\%) into a binary host consisting of the bipolar Bis[3,5,-di(9H-carbazol-9-yl) phenyl]diphenylsilane (SimCP2) and  electron transporting 2-(4-biphenylyl)-5-(4-tert-butylphenyl-1,3,4-oxadiazole) (PBD) in the emissive layer (EML) of an OLED. The device structure is as follows: ITO/PEDOT:PSS (40 nm)/PVK (15 nm)/SimCP2:PBD:Eu(DBM)$_3$Phen (30:70:5 wt.\%) (64 nm)/BCP (10 nm)/Alq$_3$ (40 nm)/LiF (0.8 nm)/Al (100 nm). In this structure, poly(3,4-ethylenedioxythiophene) polystyrene sulfonate (PEDOT:PSS), Poly (n-vinyl carbazole) (PVK), 2,9-dimethyl-4,7-diphenyl-1,10-phenonthroline (BCP), and tris(8-quinolinolato) aluminum(III) (Alq$_3$) are the hole injection, hole transporting (HTL)/electron blocking, electron transporting (ETL)/hole blocking, and electron injection layers, respectively. To investigate the magnetic field effects, a lightweight magnet with the magnitude of B = 235 mT was placed directly on the devices during the measurements. Details of the device fabrication and characterizations, the molecular structures of the materials, and the molecular energy level alignments are presented in the supplementary material (see also Fig. S1 and Fig. S2).

\section{SENSITIZATION OF THE Eu$^{3+}$ IONS IN THE EMISSIVE LAYER (EML)  }
\label{EXPERIMENTAL}

Since \emph{f-f} transitions in lanthanide ions are Laporte-forbidden, in order to sensitize them, they are usually coordinated to some organic ligands, forming so-called \emph{lanthanide complexes}.\cite{26} Upon excitation of the organic ligands in a lanthanide complex, the sensitization process can occur via F\"{o}rster-type and/or Dexter-type energy transfer mechanisms from the singlet and triplet energy levels of the organic ligands to the emissive central lanthanide ion. Such an energy transfer mechanism is called \emph{antenna effect}.\cite{26} All the possible energy transfer pathways from the singlet and triplet energy levels of SimCP2 and PBD to the triplet energy levels of the DBM and Phen ligands and subsequently to the Eu$^{3+}$ ion in Eu(DBM)$_3$Phen are shown in Fig. S3 in the supplementary material (see also Table S1). Moreover, upon incorporation of the lanthanide complex into a suitable host matrix in an OLED, the sensitization of the lanthanide ion is improved because the excitation energy of the host material can be transferred to the ligands if the absorption of the ligands overlaps with the emission from the host material.
\begin{figure}[ht]
\begin{center}
\resizebox{0.45 \textwidth}{!}{
\includegraphics{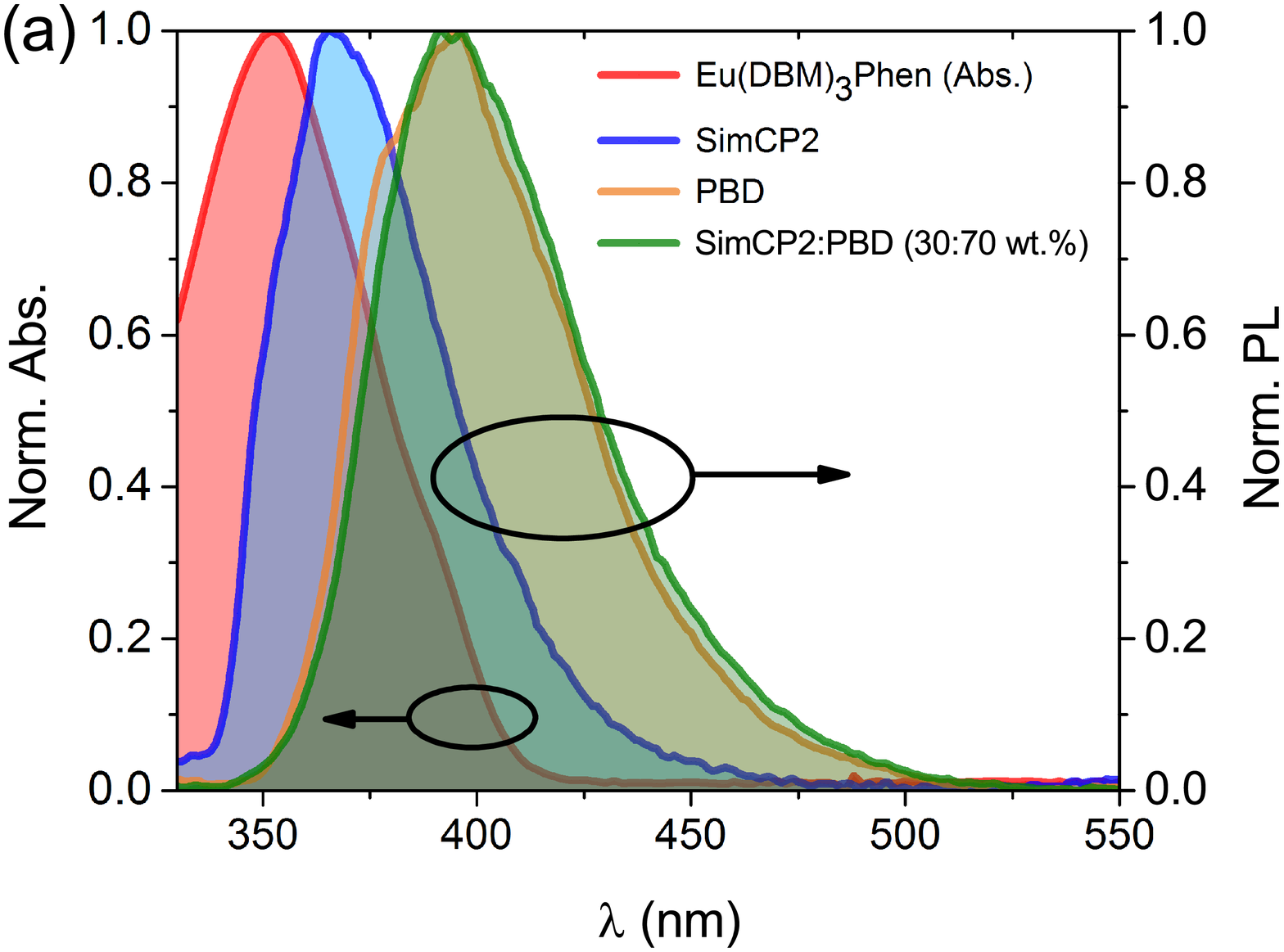}}
\resizebox{0.45 \textwidth}{!}{
\includegraphics{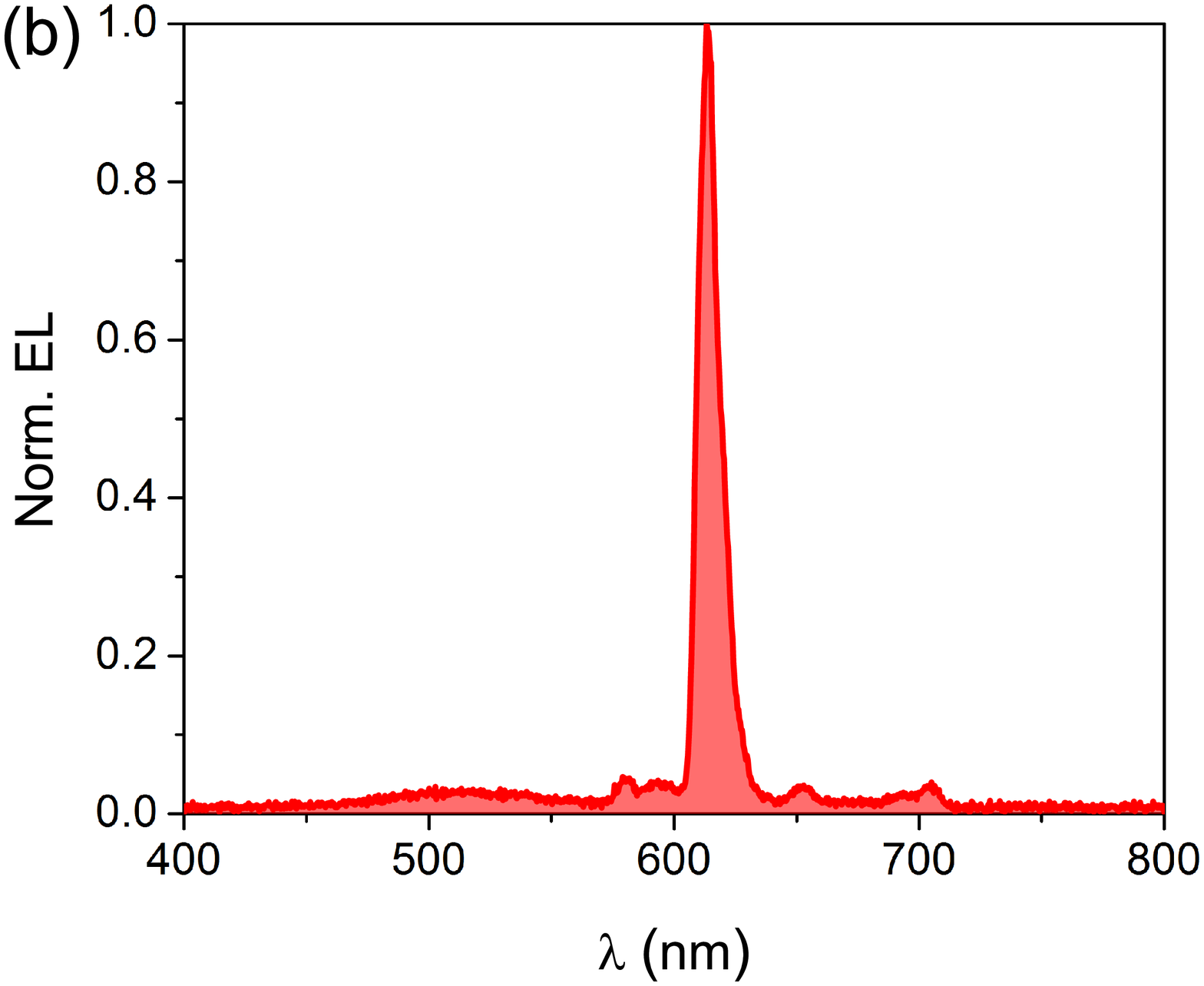}}
\end{center} \caption{(a) (Color online) The normalized photoluminescence (PL) spectra of SimCP2, PBD, SimCP2:PBD blend, and the absorption (Abs.) spectrum of Eu(DBM)3Phen, (b) the electroluminescence (EL) spectrum of the device.}
\label{fig1}
\end{figure}
As Fig.~\ref{fig1}(a) shows, the F\"{o}rster-type energy transfer from SimCP2 and PBD to Eu(DBM)$_3$Phen can be expected to be very efficient because there is a significant overlap between the absorption spectrum of the DBM ligand and the photoluminescence (PL) spectra of SimCP2 and PBD. The EL spectrum of the device is also shown in Fig.~\ref{fig1}(b). As one can  see, in addition to the main emission centered at 612 nm (corresponding to the $^5D_0$ $\rightarrow$ $^7F_2$ transition in Eu$^{3+}$) and the other less-intense emission peaks from Eu$^{3+}$, a small and broad EL emission in the 470-550 nm spectral region is also observed, which is most likely due to the exciplexes formed at the interface between PBD in the EML and PVK HTL.\cite{27} No exciplex emission in the PL spectrum of the SimCP2:PBD blend is observed. Further, the PL spectrum of the blend looks very similar to that of PBD because of the higher weight ratio of PBD in the SimCP2:PBD blend (30:70 wt.\%). Therefore, the co-host system does not contribute to the EL in the 470-550 nm spectral region. This is because the excited energy of the co-host system is fully transferred to the Eu$^{3+}$-complex.

\section{LUMINANCE ENHANCEMENT MECHANISM }
\label{LUMINANCE}
\begin{figure}[ht]
\begin{center}
\resizebox{0.4 \textwidth}{!}{
\includegraphics{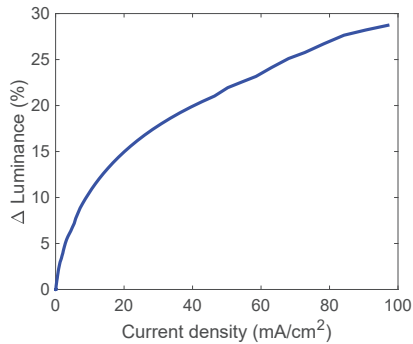}}
\end{center} \caption{(Color online) The luminance enhancement vs. current density plot under the applied magnetic field.}
\label{fig2}
\end{figure}

Since the turn-on voltage of the device was over 10 V, unfortunately, it was not technically possible in our experiments to record reliable magnetic field-dependent EL intensities at constant currents or voltages, because of the instability of the host materials. In a previous work,\cite{28} we have addressed the instability of a similar commonly-used host material in solution-processed lanthanide-based OLEDs at high voltages. For this reason, instead, we present our results for the magnetic field-dependent luminance (cd/m$^{2}$) variation -calculated from the EL spectrum and photo-current of the silicon photodiode- under quick voltage sweeps, which directly represents the EL intensity variation.  As Fig.~\ref{fig2} shows, the luminance of the device is enhanced under the external magnetic field. It is clearly seen that the luminance enhancement is even more pronounced at high current densities. Specifically, the device exhibits a luminance enhancement of up to nearly 30\% under the magnetic field, which is comparable with or even higher than the best maximum EL enhancement values (typically 10-20\%) reported for fluorescent and phosphorescent OLEDs (see, e.g., Ref. \cite{29} for a comprehensive literature review). Lanthanide-based OLEDs usually show a low level of brightness despite their excellent EL color purity.\cite{25} It is also important to note that the ultra-pure emission from visible-emitting lanthanide ions makes them irreplaceable with fluorescent and phosphorescent devices that show broad EL spectra undesired for display applications. Therefore, the enhanced luminance in our devices is promising for the realization of high-brightness lanthanide-based OLED displays.
Magnetic field-dependent luminance enhancement in lanthanide-based devices has not been studied before. We attribute the higher luminance in the presence of the magnetic field to the increased singlet (S$_0$: $|\uparrow \downarrow>$ - $|\uparrow \downarrow>$)-to-triplet (T$_{+1}$: $|\uparrow \uparrow>$ , T$_0$: $|\uparrow \downarrow>$ + $|\downarrow \uparrow>$, T$_{-1}$: $|\downarrow \downarrow>$) excited state ratio in the SimCP2:PBD co-host system, favoring the host-to-guest energy transfer via the enhanced F\"{o}rster process. More specifically, this process occurs due to the increased singlet-to-triplet e-h polaron pair ratio, resulting from the hydrogen hyperfine interactions.\cite{30} Polaron pairs can eventually evolve into their excitonic counterparts under Coulomb attraction and emit light upon e-h recombination. An external magnetic field can indeed effectively reduce the rate of intersystem crossing (ISC) between the singlet and triplet e-h polaron pairs by perturbing the electron and hole precession rates, thus increasing the theoretical 25\%-singlet exciton density limit. One should note that an external magnetic field has a minor influence on the electron and hole spin orientations in organic semiconductors.\cite{31} For example, in the present work, according to Boltzmann’s equation, exp⁡((-$\Delta$E)/kT), the magnetic field of B = 235 mT can result in only a 13 excessive spin-up per 10,000 total polarons. This is because the \emph{Zeeman Effect} ($\Delta$E) created by the external magnetic field in organic semiconductors is orders of magnitude smaller than the thermal energy (KT).\\
The triplet excited-state energy in fluorescent materials -where there is no heavy metal to favor the spin-orbit coupling- is lost because the transition from the T$_1$ to S$_0$ is forbidden in such materials. Nevertheless, the triplet energy -that is lost in the absence of external magnetic fields- can be partially converted into the singlet energy in the presence of an applied magnetic field, leading to a luminance enhancement. In that context, as the singlet-to-triplet excited state ratio in the mixed host of fluorescent SimCP2 and PBD increases under B = 235 mT, the increased singlet energy can be effectively transferred from the singlet energy levels of the host materials to the singlet energy levels of the ligands via the long-range and fast F\"{o}rster process (see Fig. S3 in the supplementary material). This energy then cascades to the triplet levels of the ligands through ISC, which subsequently excites the Eu$^{3+}$ ions, leading to a higher luminance. As shown in Fig.~\ref{fig1}(a), this is supported by the fact that there is a significant overlap between the PL spectra of the host materials and the absorption spectrum of Eu(DBM)$_3$Phen, favoring the F\"{o}rster process. Wu et al. reported a similar mechanism for the enhanced magnetic field-dependent host-to-guest energy transfer from a fluorescent host to a phosphorescent emitter in their devices.\cite{32}
\section{REDUCTION OF THE EFFICIENCY ROLL-OFF}
\label{REDUCTION}
\subsection{The exciton-exciton and exciton-polaron annihilation models}
\begin{figure}[ht]
\begin{center}
\resizebox{0.4 \textwidth}{!}{
\includegraphics{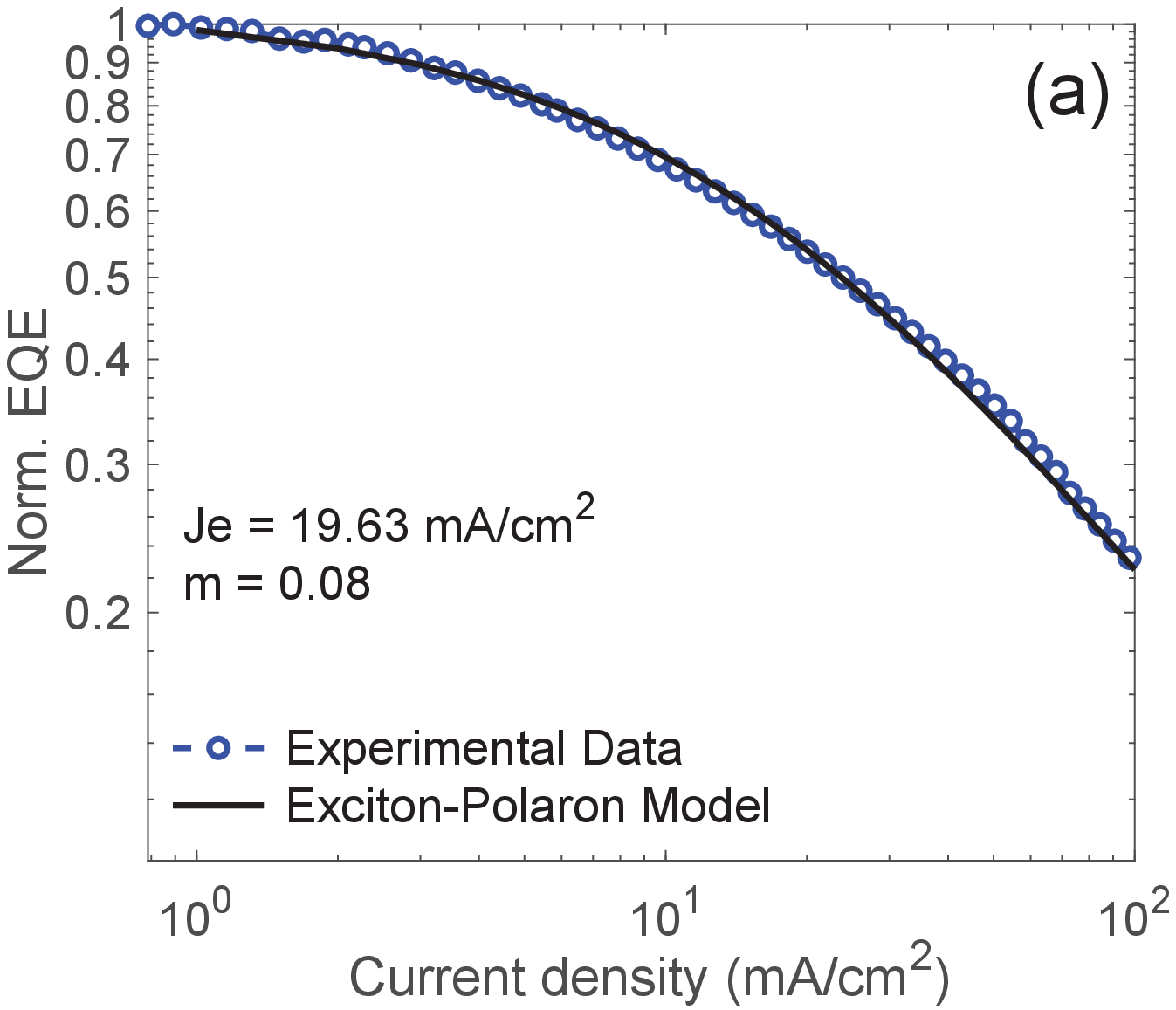}}
\resizebox{0.4 \textwidth}{!}{
\includegraphics{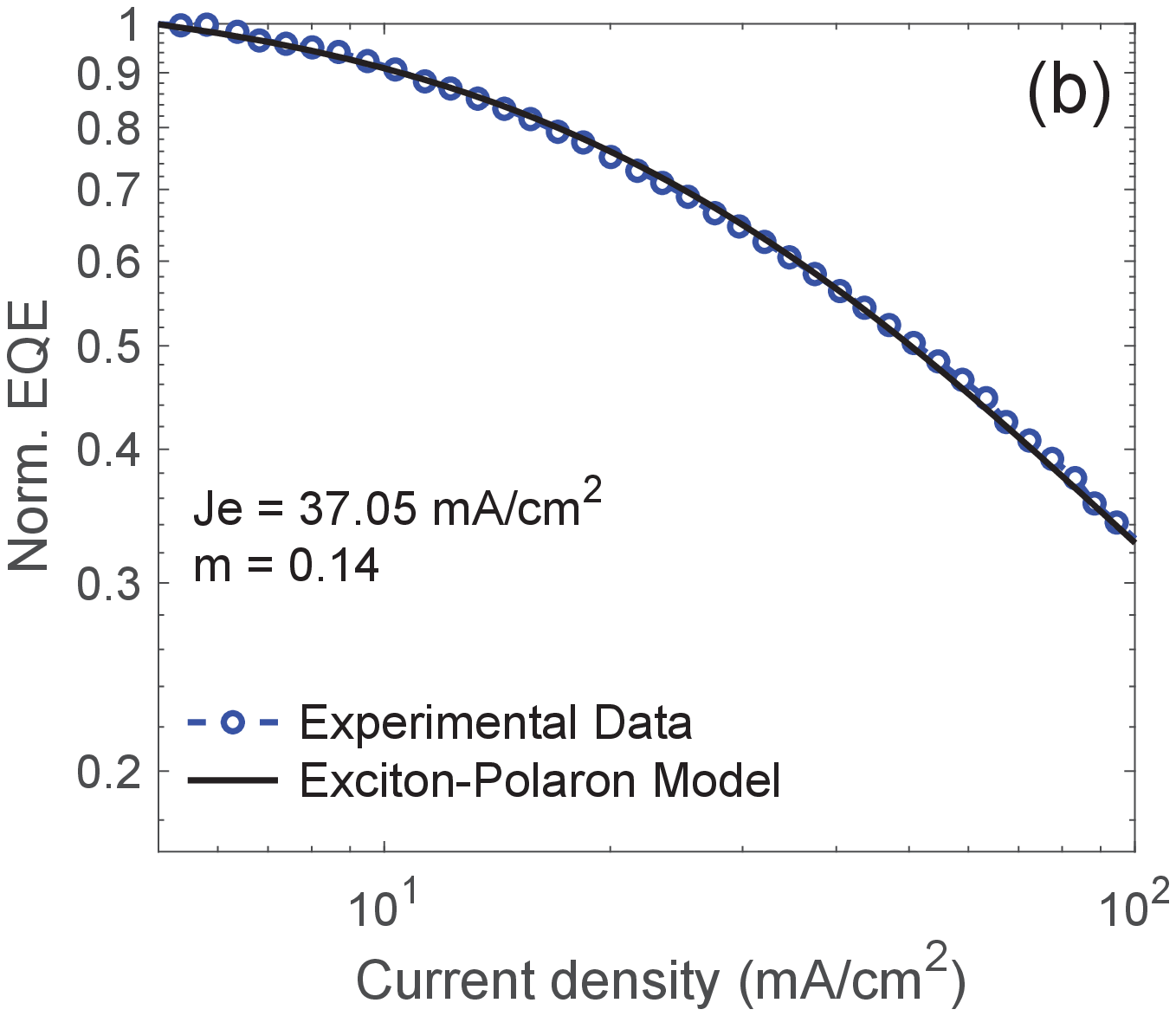}}
\end{center} \caption{(Color online) Normalized external quantum efficiency (EQE) vs. current density plots, (a) B = 0 mT, and (b) B = 235 mT, fitted to the exciton-polaron model. The fitting parameters are shown in the insets. }
\label{fig3}
\end{figure}

The normalized EQE vs. current density plots without and with the external magnetic field are shown in Fig.~\ref{fig3}(a) and (b), respectively.
We should mention here that since the maximum EQE values (occurring at very low current densities) in the presence and absence of the magnetic field in our experiments were found not to be significantly different within the uncertainty range, we analyze and discuss the results for the normalized EQE plots to provide a better clarity. As we will discuss in the following, similar to our discussion on the luminance enhancement, the effect of the applied magnetic field on the EQE roll-off  turns out to be pronounced at moderate and high current densities, at which the singlet and triplet excited state populations are expected to be large. We also show the plots separately because the current density at which the maximum EQE occurs in the presence of the magnetic field is slightly shifted compared with when no magnetic field is applied.
In Fig.~\ref{fig3}(a), without any magnetic field, the roll-off ratio that by definition is taken from the difference between the maximum EQE and the EQE at the luminance of 100 cd/m$^2$ (EQE$_{max}$ - EQE$_{100}$) is found to be 54\%. Such a severe roll-off is typically attributed to the exciton-polaron annihilation or to the exciton-exciton bimolecular annihilation mechanisms occurring between the host molecules (host-host annihilation), between the lanthanide-complex molecules (guest-guest annihilation), and between the host and guest molecules (host-guest annihilation) in lanthanide-based OLEDs.\cite{33,34,35,36,37,38,39} However, as shown in Fig.~\ref{fig3}(b) under B = 235 mT, the roll-off ratio is reduced to 34 $\pm$ 3\%, showing an improvement of 37\%.\\
Another parameter that quantifies the EQE roll-off in OLEDs is the critical current density (J$_{50}$) at which the EQE drops to 50\% of its maximum value. \cite{40} Devices with low J$_{50}$ exhibit severe roll-off behaviors. Without any magnetic field, the J$_{50}$ is found to be 23 mA/cm$^2$. However, under B = 235 mT, the J$_{50}$ increases up to 50 mA/cm$^2$. This J$_{50}$ value is higher than the reported values for the Eu$^{3+}$-based OLEDs in the literature,\cite{33,34,35,36,37,38,39} confirming the improved EQE roll-off as well in terms of J$_{50}$.\\
To further elaborate the role of exciton annihilation mechanisms in the EQE roll-off characteristics, we apply the exciton-exciton and exciton-polaron models \cite{41} to the normalized experimental EQE plots. The exciton-exciton and exciton-polaron annihilations can be characterized by the following equations:
\begin{align}
n_{XX}=\eta_0\frac{J_0}{4J}[\sqrt{1+8\frac{J}{J_0}} - 1]\nonumber \\
n_{XP}=\eta_0\frac{1}{1+(\frac{J}{J_e})^{1/(m+1)}}
\label{Eq.1}
\end{align}
where, $n_{XX}$ ($n_{XP}$) and $\eta_0$ ($\eta_0$ = 1) are respectively the EQE in the presence and absence of the exciton-exciton (exciton-polaron) annihilation processes, and J is the current density. In these equations, $J_0$, $J_e$, and $m$ are the fitting parameters.  As Fig.~\ref{fig3}(a) and (b) display, we obtain a good agreement between the experimental data and the exciton-polaron model in the presence and absence of the applied magnetic field. However, without and with the magnetic field, the exciton-exciton model results in a very poor fit (not shown) for any reasonable value of $J_0$. This indicates that the exciton-polaron mechanism is primarily responsible for the observed severe EQE roll-off. We have also previously shown that the exciton-polaron annihilation is primarily responsible for the roll-off in solution-processed lanthanide-based OLEDs.\cite{42} In this study, however, our experiments evidence that the applied magnetic field can somehow supress the exciton-polaron annihilation. To better clarify this claim, we also present some example EQE values at different current densities without and with the magnetic field in Table.~\ref{tab:1}.
\begin{table}
\caption{Summary of the EQE values at different current densities.}
\label{tab:1}       
\begin{tabular}{cccccccc}
\hline\noalign{\smallskip}
J(mA/cm$^2$)&10       &20       &30       &40       &50       &60     &70 \\
\noalign{\smallskip}\hline\noalign{\smallskip}
$EQE_0$\footnote{EQE (a.u.) at B = 0 mT.}   &$0.70$   &$0.53$   &$0.45$   &$0.40$   &$0.34$   &$0.30$   &$0.27$ \\
$EQE_{235}$\footnote{EQE (a.u.) at B = 235 mT. }
&$0.90$   &$0.75$   &$0.65$   &$0.60$   &$0.50$   &$0.45$ &$0.41$ \\
\noalign{\smallskip}\hline
\end{tabular}
\end{table}
As can be seen, in the presence of the magnetic filed, the EQE values are higher at any given moderate and high current density. In the following subsection, we provide a qualitative discussion for the mechanisms accounting for the reduced roll-off.
\subsection{Magnetic field-dependent exciton-polaron annihilation reduction}
\begin{figure}[ht]
\begin{center}
\resizebox{0.45 \textwidth}{!}{
\includegraphics{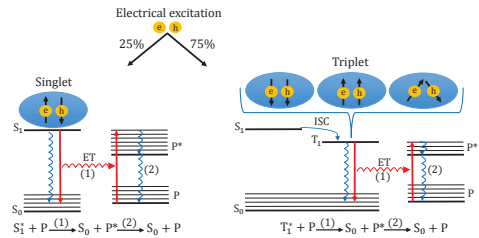}}
\end{center} \caption{(Color online) Schematic of (left) singlet-polaron and (right) triplet-polaron annihilation processes. After the electrical excitation, the singlet and triplet excited states can relax either radiatively (solid red) or non-radiatively (wavy blue), returning to the singlet ground state S$_0$. The radiative excited energy of the singlet and triplet states can be transferred via F\"{o}rster process (wavy red) to an electron or a hole polaron (P) in the ground state. This promotes the polaron to an upper excited state (P$^*$), which then relax non-radiatively back to the ground state, quenching the exciton. ISC: S$_1$ $\rightarrow$ T$_1$ intersystem crossing. ET: energy transfer.}
\label{fig4}
\end{figure}

As Fig. ~\ref{fig4} displays, the energy of an excited singlet (triplet) exciton in the S-P (T-P) annihilation process is transferred to an electron or a hole polaron via F\"{o}rster-type energy transfer mechanism, which quenches the excited singlet (triplet) exciton and consequently creates a singlet exciton in the ground state and a polaron.\cite{4,6,7}
Particularly, this mechanism is very efficient under high current densities at which the charge polaron populations at the ETL/EML and HTL/EML interfaces and the singlet and triplet exciton populations in the EML are very large. The large charge build-up at the organic/organic interfaces increases the probability for the exciton and polaron encounters. This can be evidenced when looking at the energy barriers at the ETL/EML and HTL/EML interfaces in Fig. S2 in the supplementary material. Given that the energy barrier at the ETL/EML interface in the device is much larger than at the HTL/EML interface and that electrons in organic semiconductors are much slower than holes, we believe that the observed EQE roll-off is mainly because of the annihilation of both singlet and triplet excitons mainly by the electron polarons accumulated at the ETL/EML interface. Comparing the S-P and T-P annihilation processes, we also expect that the latter is more severe in our devices because of the longer lifetime of the triplet excitons in the co-host system and also on the organic ligands, which can easily provide enough time for the quenching of the triplet excitons by the electron and hole polarons. This process should be more severe between the polarons and the host molecules because the low concentration of the guest molecules in the SimCP2:PBD:Eu(DBM)$_3$Phen (70:30:5 wt.\%) blend may not significantly contribute to the overall EQE roll-off in terms of T-P annihilation. \\
In that context, the T-P annihilation in this work is reduced because the external magnetic field decreases the triplet exciton density mainly in the co-host system, lowering the probability for the polaron interactions with the triplet excitons formed on the host molecules. One should also note that even though the reduction of triplet excitons in the co-host system in the presence of the magnetic field could make the (slow) Dexter-type energy transfer from the host materials to the ligands less efficient, the enhanced (fast) F\"{o}rster host-to-guest energy transfer mechanism overcompensates this process. This in turn should simultaneously reduce the S-P annihilation (even if it has a lesser contribution to the roll-off) on the host molecules, due to the increased singlet-to-triplet population at the moderate and high current densities, which can be supported by the enhanced luminance discussed earlier. This observation suggests that our approach for suppression of the efficiency roll-off may work even better for the devices with fluorescent emitters. As a result, the net effect of the magnetic field is the reduction of the exciton-polaron annihilation mechanisms while enhancing the luminance of the devices at high current densities.\\
Finally, as briefly mentioned in the introduction section, it has been shown that the reaction rate between triplet excitons and polarons can be reduced in the presence of an external magnetic field. In this case, triplet excitons are considered as the trapping sites for electron and hole polarons, whose interactions are interrupted in the presence of an external magnetic field (known as \emph{the site-blocking effect}).\cite{23} This effect increases the polaron mobility and thus enhances the magneto-conductivity and EL intensity. As discussed earlier, the interpretations presented for the reduced exciton-polaron annihilation in this work were based on the variation of the exciton dynamics under the applied magnetic field but our results are also consistent with the site-blocking effect which may also be happening in our devices. However, according to the literature,\cite{23,24,43,44,45,46} it requires further investigations to confirm, which is out of the focus of this paper.
\section{CONCLUSIONS} Our proof-of-concept experiments demonstrate that by modification of the excited state dynamics and excitonic processes, a relatively large external magnetic field can reduce the EQE roll-off in Eu$^{3+}$-based OLEDs, owing to the diminished exciton-polaron annihilation. The applied magnetic field also enhances the luminance of the devices. The proposed approach for reducing the roll-off in the present work is reproducible, inexpensive, and easy-to-implement into practical applications. It may also open up an avenue towards the realization of roll-off-less lanthanide-based devices with enhanced luminance. This would also be particularly important for the realization of lanthanide-based high-brightness OLED displays and electrically pumped organic semiconductor lasers.

\section*{Supplementary Material}
See supplementary material for details of the device fabrication and characterizations, the molecular structures of the materials and their molecular energy levels, and the singlet and triplet energy diagram of the host materials and the lanthanide complex.

\section*{Acknowledgments}
This wok was supported by the Natural Sciences and Engineering Research Council (NSERC), the Discovery Grants program (No. RGPIN-2015-05485), and the CREATE program (Novel Chiral Materials: An International Effort in Research and Education).

\end{document}